\documentclass[pdflatex,sn-mathphys-num]{sn-jnl}

\usepackage{graphicx}%
\usepackage{multirow}%
\usepackage{amsmath,amssymb,amsfonts}%
\usepackage{amsthm}%
\usepackage{mathrsfs}%
\usepackage[title]{appendix}%
\usepackage{xcolor}%
\usepackage{textcomp}%
\usepackage{manyfoot}%
\usepackage{booktabs}%
\usepackage{algorithm}%
\usepackage{algorithmicx}%
\usepackage{algpseudocode}%
\usepackage{listings}%

\theoremstyle{thmstyleone}%
\theoremstyle{thmstyletwo}%

\theoremstyle{thmstylethree}%

\begin{document}

\title[Article Title]{Movie Recommendation with Poster Attention via Multi-modal Transformer Feature Fusion}


\author[1]{\fnm{Linhan} \sur{Xia}}
\author[1]{\fnm{Yicheng} \sur{Yang}}
\author[1]{\fnm{Ziou} \sur{Chen}}
\author[1]{\fnm{Zheng} \sur{Yang}}
\author*[2]{\fnm{Shengxin} \sur{Zhu}}\email{Shengxin.Zhu@bnu.edu.cn}

\affil*[1]{\orgdiv{ Guangdong Provincial Key Laboratory of Interdisciplinary Research and Application for Data Science}, \orgname{BNU-HKBU United International
College}, \orgaddress{ \city{Zhuhai}, \postcode{519087}, \state{Guangdong}, \country{China}}}

\affil[2]{\orgdiv{Research Center for Mathematics}, \orgname{Beijing Normal University}, \orgaddress{\street{Jingfeng Road}, \city{Zhuhai}, \postcode{519087}, \state{Guangdong}, \country{China}}}


\abstract{Pre-trained models learn general representations from large datsets which can be fine-turned for specific tasks to significantly reduce training time. Pre-trained models like generative pretrained transformers (GPT), bidirectional encoder representations from transformers (BERT), vision transfomers (ViT) have become a cornerstone of current research in machine learning. This study proposes a multi-modal movie recommendation system by extract features of the well designed posters for each movie and the narrative text description of the movie. This system uses the BERT model to extract the information of text modality, the ViT model applied to extract the information of poster/image modality, and the Transformer architecture for feature fusion of all modalities to predict users' preference. The integration of pre-trained foundational models with some smaller data sets in downstream applications  capture multi-modal content features in a more comprehensive manner, thereby providing more accurate recommendations. The efficiency of the proof-of-concept model is verified by the standard benchmark problem the MovieLens 100K and 1M datasets. The prediction accuracy of user ratings is enhanced in comparison to the baseline algorithm, thereby demonstrating the potential of this cross-modal algorithm to be applied for movie or video recommendation.}




\maketitle

\section{Introduction}\label{sec1}

The development and application of information technology has generated overload information \cite{b1} and it have become a difficult task for users to retrieve relevant information that matches their needs from the overload information. It may even cause a decrease in the quality of users' decision-making as well as an increase in their anxiety level \cite{b2}.

In order to help users better filter information through reasonable information management and filtering mechanisms, recommender systems have emerged. The role of a recommender systems is to filter information that may interest users based on their interests, behaviors and preferences, and to provide personalized content in the massive amount of information to enhance user experience and satisfaction \cite{b3}.

Recommender systems usually integrate data mining, machine learning and other means to analyze users’ historical behavior, social relationships and content characteristics of the information to build user-item interaction model \cite{b4}. The advent of recommender systems in the mid-1990s saw the emergence of algorithms that matched and customized content for users. These systems have since become a pervasive feature of e-commerce, social media, audio-visual entertainment, and numerous other fields.

The development of deep learning in recent years has greatly empowered recommender systems, and some of these methods are already working in real applications. For instance, YouTube uses Convolution Neural Networks(CNN) to extract poster features combined with user preferences \cite{daneshvar}; Netflix uses recurrent neural networks (RNN) to track users' reading history to capture changes in user interests \cite{amatriain}; twitter uses graph neural networks (GNN) to analyze users‘ social relationships and interaction history to achieve personalized topic recommendations \cite{li}. What's more, recommendation systems can also be used for medical care \cite{zheng2024} and knowledge discovery \cite{chen2020}.

However, with the popularization of internet applications and the proliferation of multimedia contents, it is difficult to meet users' needs with single-modal information. Among them, the data sparsity problem is a serious challenge. In the context of a recommendation system, the term 'sparsity problem' is used to describe a situation where interaction data between users and items is very sparse. This makes it challenging for the system to accurately learn and predict users' preferences. Cross-modal recommender systems are supposed to capture content features more comprehensively by integrating and analyzing data from multiple modalities and thus provide more accurate recommendations. For instance, in movie recommendation, the data sparsity problem is alleviated by combining information such as text descriptions and poster images of movies.

This research proposes a movie recommender system that combines cross-modal data analysis and deep learning. The proposed system employs pre-trained model to mine and fuse multi-source information and uses a transformer architecture to fuse them. The multi-modal data set includes user characteristics and multi-modal data of movies. Result on the MovieLens benchmark problems shows that better prediction accuracy can be achieved in comparison to some traditional algorithms. This indicates the potential of multi-model recommendation. 

\section{Related Work}\label{sec2}

In the past two decades, the development of recommendation algorithms has had a profound impact on people's lives. Major content platforms such as Bytedance, Weibo, and Twitter have adopted different recommendation algorithms to serve their users and help them discover personalized content. With the development of the Internet industry, numerous researchers have put forward a variety of different methods. This chapter describes the relevant work in detail from two aspects: traditional approaches and multi-modal fusion approaches. 

\subsection{Single modal approaches of recommendation system}

Single modal recommendation algorithms can be divided into three main branches: content based approach \cite{b5}, collaborative recommendation approach \cite{b6} and hybrid recommender system \cite{b7}.

Content-based recommender systems have evolved over decades as well as relatively matured, and their core features are relying on content features and user characteristics to achieve accurate recommendations. Roy et al. provide an extensive overview of these principles, and they emphasize the importance of understanding user preferences and content features to generate accurate recommendations \cite{r1}. The process of understanding user preference is often referred to as profile inference \cite{lu2019}. In addition, Dahdouh et al. emphasized the significance of tools for recommender system development, with particular reference to big data technologies such as Hadoop and Spark in content-based recommender systems \cite{dahdouh2019large}.

Content based recommender systems rely on content features to generate recommendations. Based on the traditional algorithmic perspective, some scholars have proposed to analyze the text images or other content forms of the items to match with user preferences. Beel et al. have used the TF-IDF approach to capture the semantic relevance of the documents to achieve accurate recommendations in the publication domain \cite{r3}. Latent semantic analysis (LSA) is a modeling technique based on singular value decomposition, which aims to reduce the dimensionality of document features and discover the potential topics in them. Bergamaschi et al. used LSA to model the topics of the documents, which significantly improves the effectiveness of the recommender system \cite{r4}. In addition, rule-based recommender system also favored by some researchers. Bouihi et al. proposed a recommendation algorithm based on learning history rule and learning performance rule for recommendation based on embedded coding \cite{r5}. Simple Bayesian classifier is a probabilistic based classification algorithm commonly used in text classification and recommender systems with the advantages of computational efficiency and simplicity of implementation. Roy et al. in their study pointed out the application of simple Bayes in recommender systems especially in news recommendation and email filtering \cite{r1}. Zhang et al. proposed a method for linearly classifying text content using SVMs and thus realizing accurate recommendation approach, emphasizing its role in improving accuracy and generalization ability \cite{r6}. Gao et al. proposed to utilize a linear mixed model for group recommender systems, based on this approach the accuracy and adaptability of the recommendation system in group recommendation are improved effectively \cite{gao2019}. Zuo et al. proposed a approach to combine linear mixed model and singular value decomposition (SVD) to improve the performance \cite{zuo2020}.

The development of deep learning has provided new perspectives for content based recommender systems. Shafqat et al. employed hierarchical LSTM model to predict the  destination of travelers. This approach allowed the model to learn the context, thereby enabling it to consider factors such as location, user, and environment. These factors included aspects such as weather, climate, and risk \cite{r7}. Paradarami et al.  proposed a deep learning neural network framework that utilizes reviews and content-based features to generate model-based predictions for business user portfolios \cite{r7}. Hassan et al. proposed the use of artificial neural networks to model standard ratings and determine the predictive performance of the system based on the aggregated function approach \cite{r9}. The performance is more prominent in seven evaluation metrics. David et al. evaluated the performance of various models including LSTM, GRU in e-commerce recommender systems and found that RNN models based on the attention mechanism have certain advantages on specific tasks \cite{r10}. In order to achieve holistic optimization of long-term recommendation, Huang et al. proposed a reinforcement learning-based long-term recommendation system. The recommendation process in Huang's work is considered as a Markov decision process, in which the interaction between agents (recommender system) and environments (user) is simulated by a RNN, and a reinforcement learning optimization model is used to achieve more accurate recommendation \cite{r11}. Lee et al. and others divided the recommendation cycle into different time steps and the proposed recommendation system based on RNN divides the time series into multiple periods for recommendation \cite{r12}. These deep learning models can capture higher order feature interactions \cite{xu2021} and explore users deep interests \cite{feng2022,luo2023} and thus often result in better results. Siet et al. proposed a movie recommendation system, which used user behavior sequence and multi-head attention mechanism to analyze user preferences, embed user information and movie sequences to optimize data processing results, and finally use the K-means algorithm to improve the accuracy and diversity of recommendations \cite{Siet}. Aljunid et al. used deep learning to embed the characteristics of users and movies, and achieves dimensionality reduction through embedding layer, and predicts the score through forward propagation method \cite{Aljunid}.

A collaborative recommendation system is a type of recommendation technology that employs user behavior data, which is widely used in e-commerce, social media, movie recommendation and other domains. In recent years, with the expansion of data and the enhancement of algorithms, the research of collaborative recommendation systems has witnessed a significant advancement.

One of approaches is user-based collaborative filtering, which identifies groups of users with similar interests to the target user and formulates recommendations based on their preferences. Ma et al. improved the user-based collaborative filtering (CF) technique by proposing two recommendation methods that incorporate user-generated labels and social relationships. The performance is improved compared to the baseline model \cite{r13}. Millan et al. used KNN algorithm to generate asymmetric group, based on the asymmetric group to measure the asymmetric similarity among users and thus achieve recommendation \cite{c1}. O'connor et al. proposed PolyLens, which is designed to provide recommendation services for groups of users rather than for individuals to recommendation, and they pointed out that giving up some privacy in group recommendation can get better recommendation results \cite{c2}. Zhang et al. proposed to extract the hidden social relationship information based on user feedback, and integrate the user's social relationship information into matrix factorization and Bayesian personalized ranking to solve the problems of rating prediction and item ranking, respectively \cite{c3}. Fakhri proposed a methodology based on the user rating similarity and user similarity for restaurant recommendation, which has achieved an mean absolute error (MAE) of 1.492 in an experiment without user attributes \cite{c4}.

Deep learning algorithms also have a wide range of applications in collaborative recommender systems. Fu et al. proposed to simulate effective intelligent recommendations by implementing learning about users and content \cite{c5}. Fu's approach embeds encoded user-items relevance semantic information, respectively, to obtain low latitude vectors as a way of realizing the recommendations. Bobadilla et al. proposed a three-phase approach to collaborative recommendation \cite{c6}. The first phase involves the actual prediction error, the second the prediction error, and the third the prediction rating. Each phase has its own learning process, which aims to improve the quality of prediction at a given time. Wang et al. Collaborative deep learning, which inherits the idea of collaborative topic regression, couples two different sources of information together, and generates rating matrices for collaborative filtering through deep learning \cite{c7}. Aljunid et al. proposed a multi-model deep learning (MMDL), integrating items and users to build a hybrid recommender system, MMDL combines deep self-encoder and one-dimensional convolution to realize the improvement of performance \cite{c8}. The research of Wu \cite{c9} pointed out that there is a great potential of graphical neural networks (GNNs) to be applied in recommender systems. Based on GNN, Gao et al. proposed CA-GNN for sensing contextual scenes and users-items interaction graphs, and used the attention mechanism to improve the interpretability and accuracy of the model \cite{c8}. Wang et al. proposed a graph neural network social recommendation model, for GNN-based social recommendation \cite{c11}. 

Hybrid recommender systems integrate multiple recommendation techniques, capitalising on the unique strengths of each to enhance the overall efficacy of the system. These systems frequently employ collaborative filtering, content-based recommendations, and other algorithms, such as machine learning, to enhance the precision and efficacy of the recommendations generated. Kreutz et al. combined content-based and collaborative filtering techniques to provide personalized recommendations for researchers, generating personalized recommendations based on users' historical publications and their citation patterns \cite{kreutz}. Pradhan et al. proposed a paper recommendation combining social network analysis and content-based recommendation, the contextual similarity of papers and social relationships between researchers are considered \cite{pradhan}. Lorbeer et al. employed deep learning techniques to analyze textual data and clustering algorithms such as birch to improve the grouping of similar items and enhance the content-based filtering process \cite{lorbeer}. Qaiser et al. proposed an article recommendation algorithm based on word embeddings generated from article metadata to form article embeddings, followed by spherical k-means clustering, combined with content-based methods and clustering to improve scalability \cite{qaiser}. Biswas et al. combined collaborative filtering and deep learning to solve the cold start problem and improve the recommendation quality by using user behavior data and item characteristics \cite{biswas}.

\subsection{Multi-modal fusion recommendation system}

The advent of multimedia platforms has prompted a surge of interest among researchers in recommender systems based on multi-modality. These systems are able to consider multi-modal data (such as input text and image) simultaneously, offering a more comprehensive approach to information processing.

In the level of multi-modal recommendation algorithms, Saga et al. proposed a way to extract features of product images based on convolutional neural networks (CNNs), combined with information from other modalities such as user reviews to achieve prediction of users' purchasing tendencies \cite{c12}. Choudhury et al. proposed a trust matrix approach combining user similarity with weighted trust propagation in multi-modality, where active users are evaluated by different trust filters and cold users generate optimized score predictions through their preferences \cite{m1}. Nikzad-Khasmakhi et al. proposed Berters to recommend relevant experts on specific topics based on three different scores: authority, text similarity and reputation \cite{m2}. Chung et al. proposed a multi-modal collaborative recommendation algorithm based on the attention mechanism, which firstly utilizes a convolutional network to represent the image features at a high order, and the user habits and preferences learned by the LSTM to perform the feature fusion to realize the recommendation algorithm \cite{m3}. In order to better embed multi-modal information, Sun et al. proposed Multi-modal Knowledge Graph Attention Network (MKGAT), which first propagates the multi-modal knowledge graph for knowledge propagation, and then aggregates the representations of different modalities for recommendation in order to improve the recommendation quality \cite{m4}.

The core of multi-modal recommendation algorithm is how to fuse the information of different modes. By fusing the information of multiple modes together, complementary information can be provided for the model to improve the accuracy of the model \cite{m5}. For this reason, some scholars proposed different fusion methods.

Linear weight based fusion method is the simplest fusion approach which combines information from different modalities in a linear manner. Foresti and Snidaro and Yang et al. applied this approach to detecting and tracking pedestrians \cite{F1,F2}. Zhu et al. proposed a low-rank tensor multi-modal fusion approach with an attentional mechanism, which improves the efficiency and reduces the computational complexity at the same time \cite{F3}. Wei et al. proposed the use of cross-attention architecture to achieve fusion of image modalities and text modalities for downstream tasks \cite{F4}. Wang et al. proposed a multi-modal token fusion method for transformer-based multi-modal token fusion (TokenFusion), where TokenFusion dynamically detects uninformative tokens and replaces these tokens with projected and aggregated multi-modal features \cite{F5}. This design helps the model learn the correlation between different modalities.

\subsection{Pre-trained models}

In our research, we implement recommendation algorithm based on feature extraction through pre-trained models. Pre-trained models have become a cornerstone of current research in machine learning, especially in the fields of Natural Language Processing (NLP) and Computer Vision (CV). Pre-training models learn general representations from large datasets, which can be fine-tuned for specific tasks to significantly reduce training time.

In the field of NLP, pre-trained models such as BERT (Bidirectional Encoder Representations from transformers) \cite{devlin}, Generative Pre-trained Transformer (GPT) \cite{P1} and Robustly optimized BERT approach (RoBERTa) \cite{liu2021robustly}. They are pre-trained in a corpus, initially learn the knowledge of human society and embed it in their own parameters and can be fine-tuned in downstream tasks to achieve the target results.

In the field of CV, pre-trained models such as VGGNet \cite{P3}, Residual networks (RoBERTa) \cite{P4} and Vision transformer (ViT) \cite{P5} based on transformer architecture. These models are pre-trained on large-scale image datasets to learn to extract image features and migrate these features to other tasks including feature extraction, target detection and semantic segmentation.

\section{Methodology}\label{sec4}
In this section, we elaborate the composition and structure of our cross-modal movie recommendation algorithm based on pre-trained model feature extraction and Transformer architecture feature fusion. The model uses BERT model to extract the information of text modality, the ViT model to extract the information of image modality, and the Transformer architecture for feature fusion of all modalities to predict the users' ratings of movies. The performance indexes are significantly improved compared with traditional collaborative filtering algorithms.

\subsection{Framework of the proposed model}

\begin{figure}[H]

\centering
\includegraphics[width=\textwidth]{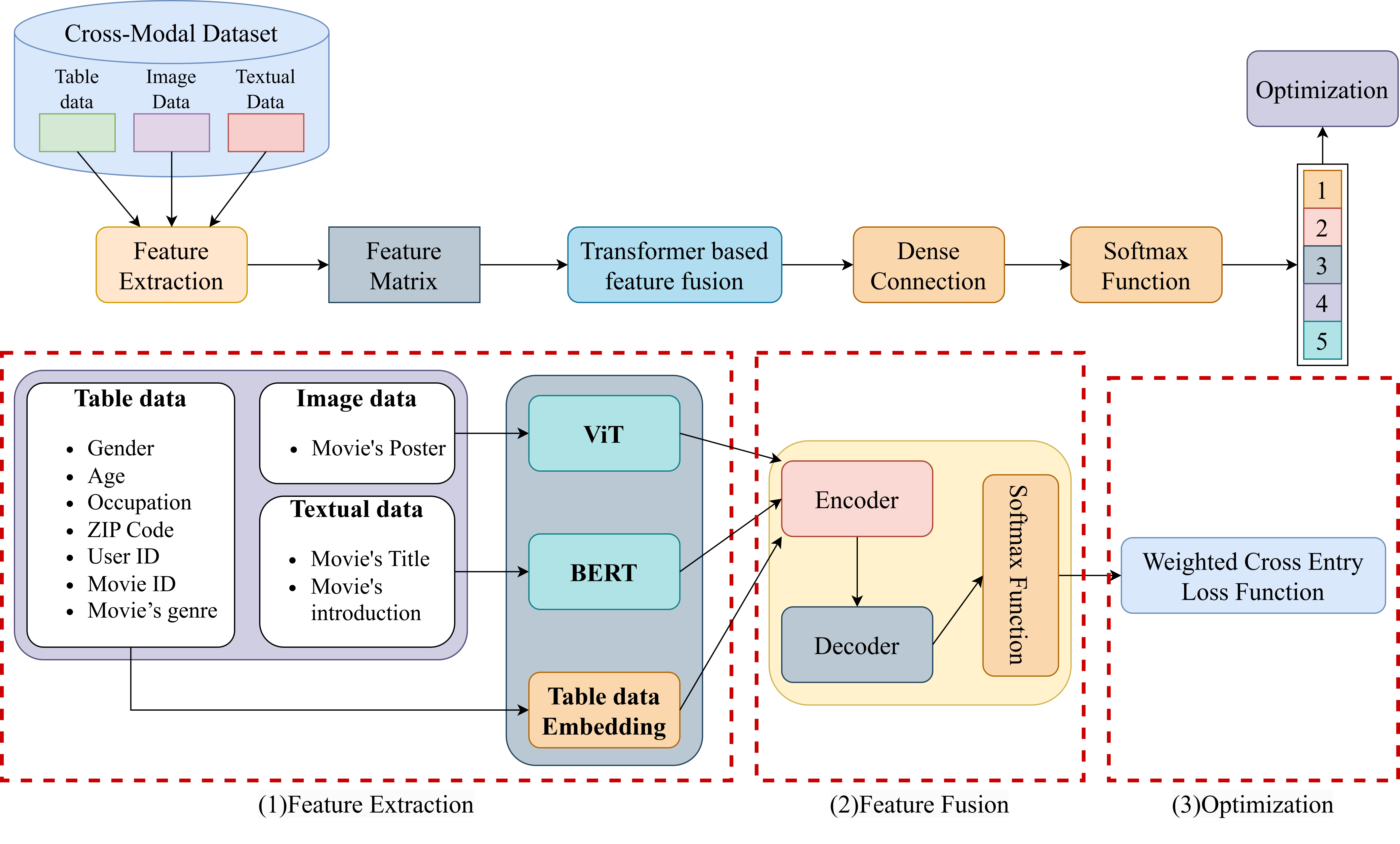}

\caption{Pipeline of proposed model where ViT is Vision transformer and BERT is Bidirectional Encoder Representations from transformer. Both of them are pre-trained model downloaded from Hugging Face.\label{Pipeline}}
\end{figure}

We propose a deep learning based cross-modal neural network for movie recommendation. The framework of our model is illustrated in Figure \ref{Pipeline}. The input data of the model comprises multi-modal information from both users and movies. Users' information includes the user ID, age, ZIP code, occupation, and gender, while movies' information comprises the movie ID, genre, poster, title, and introduction. These data are divided into three modalities: textual information modality (e.g., title and introduction), image information modality (e.g., movie poster) and structured data (e.g., genre, occupation, etc.). We use different embedding methods to embed information in each of the three modalities. For image modality we use ViT model to extract the features, for text modality we use BERT model to extract the features, and for structured data we design a multilayer neural network based embedding structure to embed the information. After feature extraction, we design a transformer-based feature fusion module, and input the fusion result into the classification function to get the user's rating estimation of a certain movie so as to realize personalized recommendation.

\subsection{Textual feature extractor}

\begin{figure}[htbp]
\centering
\includegraphics[width=0.7\textwidth]{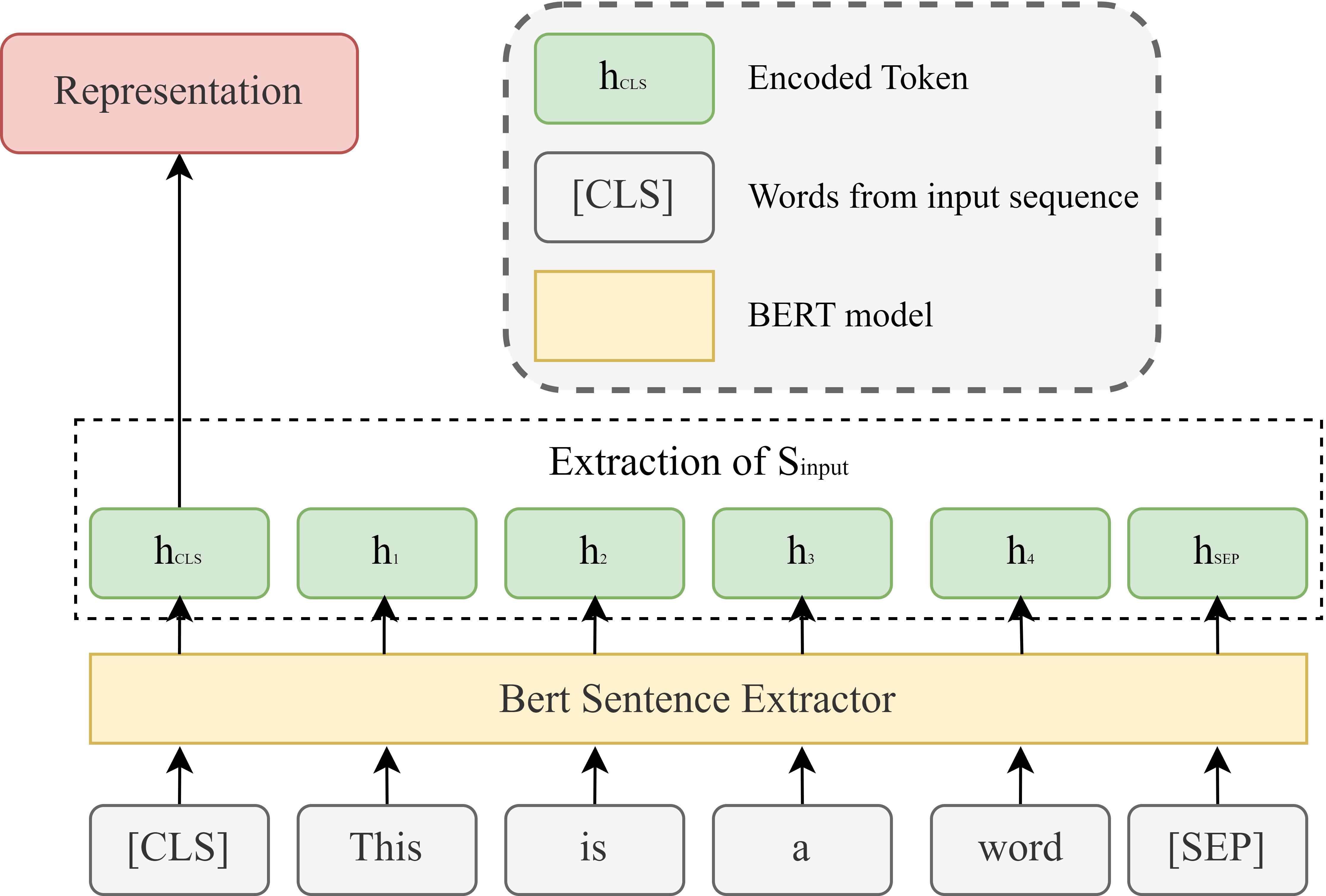}
\caption{Feature extraction process of BERT model in our research, the dimension of BERT's out put is 768.}
\label{fig:bert-extraction}
\end{figure}

For the information of text modality we use BERT model for feature extraction. We denote the input text sequence as \(S_{\text{input}}=[W_1,W_2,\ldots,W_{\text{len}}]\), where \(W_x (1 \leq x \leq \text{len})\) is a word (later denoted as Token) in the input text sequence. The BERT model firstly adds "[CLS]" token (denoted as \(W_{\text{CLS}}\)) and "[SEP]" token (denoted as \(W_{\text{SEP}}\)) before and after the text sequence as the start and end of the sequence respectively. The BERT model encodes the sequence to obtain a hidden state matrix \(H=[h_{\text{CLS}},h_1,h_2,\ldots,h_{\text{len}},h_{\text{SEP}}]\) where \(h_{\text{CLS}}\) and \(h_{\text{SEP}}\) are the hidden value of \(h_{\text{CLS}}\) and \(h_{\text{SEP}}\). For the matrix \(H\), its dimension is \([768,\text{len}+2]\), where 768 is the dimension of the encoded output of the BERT model for each Token. We extract \(h_{\text{CLS}}\) as the representation of the whole sequence semantics, its dimension is 768 in our research. Its flow is shown in Figure \ref{fig:bert-extraction}.

\begin{enumerate}
    \item \textbf{Global context integration: }the first token in BERT model interacts attentively with other tokens in the whole sequence during the process of multi-head self-attention computation, so that it can learn the global information of the whole sequence.
    \item \textbf{Special pre-trained task: }the pre-trained task of the BERT model consists of two stages, one is the Masked Language Model (MLM) and the other is the Next Sentence Prediction (NSP) \cite{devlin}. In the NSP task, the BERT model needs to determine whether two sentences are contextual or not, and this process requires \(h_{CLS}\) as the semantic representation for downstream judgment, which makes the BERT model give \(h_{CLS}\) the function of learning the semantics of the whole sequence during the pre-training process.
\end{enumerate}

Through these mechanisms, \(h_{CLS}\) can capture the global semantic information of the input text and provide strong support in a variety of downstream tasks.

\subsection{Image feature extractor}

For image modal features, we perform feature extraction using Vision Transformer (ViT), model that utilizes the Transformer architecture for image feature extraction. In contrast to traditional convolutional neural networks (e.g., VGGNet and ResNet), the input to ViT is not a complete matrix of pixels, but rather the image is sliced into fixed-size patches, each of which is flattened and mapped to a high-dimensional space by a linear link to form a vector representation. Assuming that the original length and width of an image are \(H\) and \(W\) respectively, the image will be cut into \(P\times P\) patches, each of which has a size of \(\frac{H\times W}{P^2}\). 

The calculation process of linear transform is shown as:

\begin{equation}
x_p^i = \text{Linear}\left(x_{\text{patch}}^i\right),
\label{eq:linear_rep}
\end{equation}where \(x_{\text{patch}}^i\) is \(i\)th patch and \(x_p^i\) is the corresponding linear representation after transformation.

Similar to the BERT model, ViT processes these patch vectors by adding a special classification token [CLS] in front of the sequence of these patch vectors, and the initial value of this token is a vector that can be learned during the training process. The whole input sequence can be represented as:

\begin{equation}
z_0 = \left[x_{\text{CLS}}; x_p^1; x_p^2; \ldots; x_p^N\right] + E_{\text{pos}},
\label{eq:z0}
\end{equation}where \(E_{\text{pos}}\) is position code, which is used to retain the position information of the patch from the image. 

After going through multiple layers of Transformer Encoder, the [CLS] token will aggregate the information of all patches and become a global representation of the entire image semantics. Eventually, the output representation of the [CLS] token is used as the semantic embedding representation of the image, which can be used for various downstream tasks, such as image classification, target detection. The formula reads as:

\begin{equation}
h_{\text{CLS}}^{L} = \text{Transformer Layers}(z_0),
\label{eq:CLS}
\end{equation}where \(h_{\text{CLS}}^{L}\) is the semantic representation of whole image.

There are several advantages of using \(h_{\text{CLS}}^{L}\) as semantic embedding:

\begin{enumerate}
    \item \textbf{Global information aggregation: }since the \(h_{\text{CLS}}^{L}\) interacts with all patches in each layer of the Transformer, it is able to capture global information about the entire image, not just local features.
    \item \textbf{Simplified structure: }compared to traditional convolutional neural networks that require the full representation of the image, the direct use of ViT's \(h_{\text{CLS}}^{L}\) simplifies the subsequent task implementation.
\end{enumerate}

\begin{figure}[H]
\centering
\includegraphics[width=\textwidth]{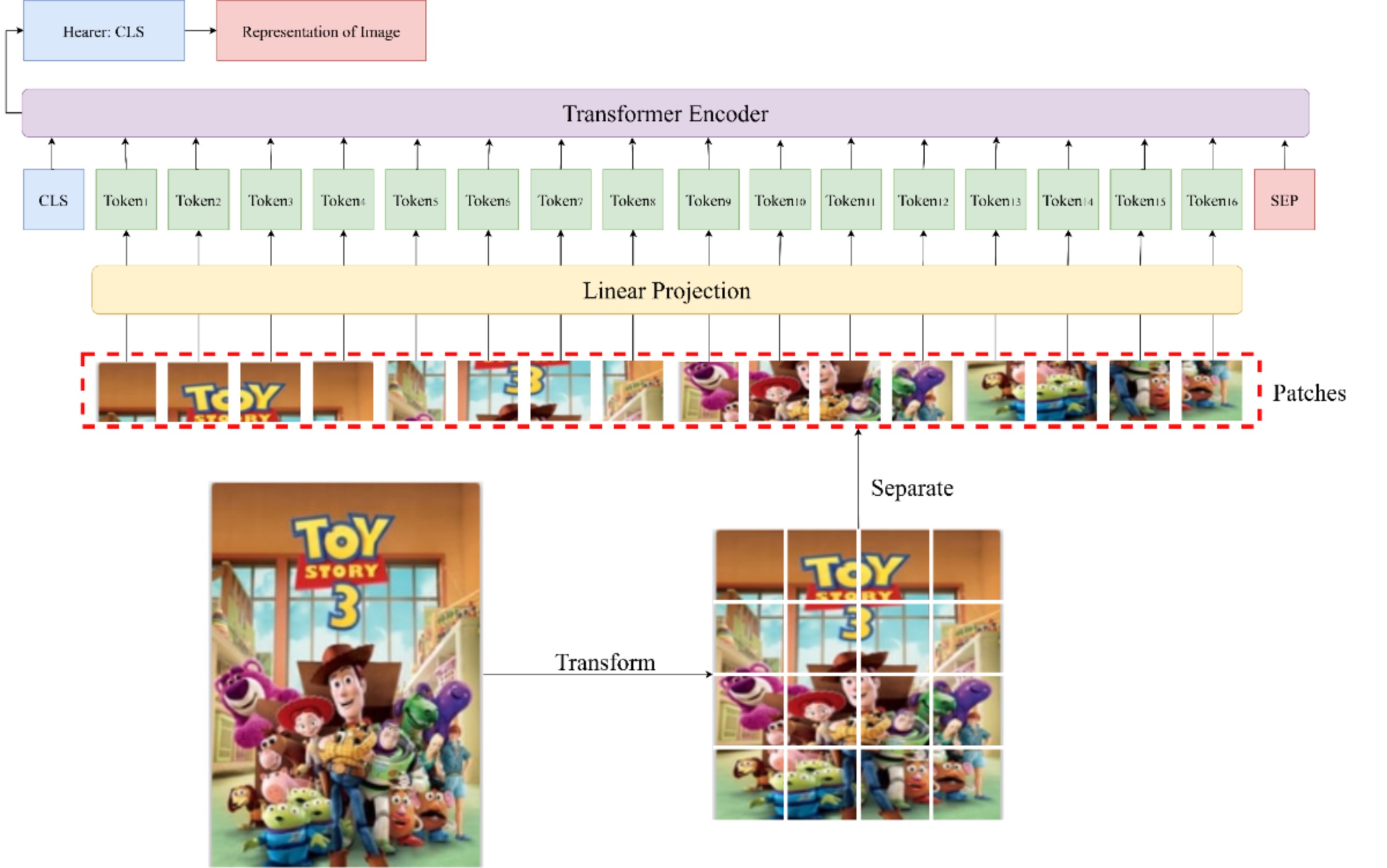}
\caption{Working process of proposed image feature extraction method,Where transformation is converting all images to a standard 224 by 224 square.\label{fig:ViT}}
\end{figure}

The work process of ViT in our research is shown in Figure \ref{fig:ViT}.

\subsection{Structural data feature extractor}

To handle structured features such as user ID, movie ID, ZIP code, etc., we use a dense layer for embedding encoding. This approach projects these discrete features into a continuous high-dimensional vector space. For this purpose we adopt an up-sampling method to embed the low dimensional features into a 768 dimensional high latitude vector space step by step, firstly for each structured feature we use different low dimensional coding methods such as one-hot encoding for the occupation feature, and then the low dimensional vectors are up-sampled by the dense layer until their dimension reaches 768. In the proposed method the ReLU function is applied as activation function.

\subsection{Feature fusion based on Transformer}

After feature extraction, we need to carry out feature fusion of multi-source features to make full use of their complementary information. To this end, a feature fusion method based on Transformer is proposed to capture the global dependencies between different modal features through self-attention mechanism to achieve fusion.

In the proposed method, we first reconstruct different feature vectors into a unified feature sequence, and input each feature vector into transformer as a Token. We add special tokens before and after the sequence: [CLS] and [SEP], the former to mark the semantic information of the whole sequence, and the latter to mark the end of the sequence.

The constructed feature sequence is denoted as \(S=[V_{CLS},V_1,V_2,...,V_{10},V_{SEP}]\). Where \(V_i(1\leq i\leq 10)\) is \(ith\) feature vector in the sequence, \(V_{CLS}\) and \(V_{SEP}\) are special tokens. 

Since a transformer is a model without sequential information, in order to enable the model to distinguish different feature tokens, we use fixed-position coding to add location information to feature vectors. Position coding is a vector with the same input feature dimension, that is, 768 dimensions. By introducing position coding model, the position of each feature vector in the sequence can be identified and sequence information can be captured. The calculation formula of location encoding reads as

\begin{equation}
P_{(pos,2i)}=\sin{\left(\frac{pos}{10000^{\frac{2i}{d}}}\right)},
\label{eq:pos1}
\end{equation}

\begin{equation}
P_{(pos,2i+1)}=\cos{\left(\frac{pos}{10000^{\frac{2i}{d}}}\right)},
\label{eq:pos2}
\end{equation}where \(pos\) represents the position, \(i\) represents the dimension. In practice, we add the positional coding matrix \(P\) to a feature sequence \(V\) to get the positional encoded feature sequence, the process reads as:

\begin{equation}
    V_{i}^{pos}=V_i+P.
    \label{eq:feature}
\end{equation}

The position encoded feature sequence is used as input for the transformer encoder. The transformer encoder consists of multiple self-attention layers and feedforward neural network layers that capture global associations between features through multi-head self-attention mechanisms.

At the heart of the self-attention mechanism lies the calculation of the importance of each token to the other tokens in the input sequence, which is achieved through three matrices: Query (Q), Key (K), and Value (V). These matrices are obtained from the input features by linear transformations as:

\begin{equation}
    Q=V_{i}^{pos}\cdot W_Q,
    \label{eq:transformer-q}
\end{equation}

\begin{equation}
    K=V_{i}^{pos}\cdot W_K,
    \label{eq:transformer-k}
\end{equation}

\begin{equation}
    V=V_{i}^{pos}\cdot W_V,
    \label{eq:transformer-v}
\end{equation}where, \(W_Q,W_K,W_V\) are trainable weight matrix. After we get \(Q,K,V\) we can conduct self attention calculation as:

\begin{equation}
    \text{Attention}(Q,K,V)=\text{SoftMax}\left(\frac{Q\cdot K^T}{\sqrt{d_k}}\right)\cdot V,
    \label{eq:attention}
\end{equation}where \(d_k\) is dimension of key. With this calculation, the model is able to assign a weight to each feature vector, thus capturing the relationship between the features.

The multi-head self-attention mechanism further parallelizes the above processes and enhances the expressive power of the model. The process reads as:

\begin{equation}
    \text{MultiHead}\left(Q,K,V\right)=\text{Concat}(head_1,...,head_h)\cdot W_o,
    \label{eq:multi-head}
\end{equation}where, \(head_i=Attention(Q,K,V)\), \(h\) is number of heads and \(W_o\) is weight matrix of output layer.After several self-attention layers and feed forward neural network layers, we get the output matrix of Transformer encoder as 

\begin{equation}
    H=\text{Transformer Encoder}\left(V_i^p\right),
    \label{eq:transformer-encoder}
\end{equation}where, \(\mathbf{H} \in \mathbb{R}^{12 \times 768}\) is output of transformer encoder. Base on the output, we extract [CLS] token as representation of whole sequence \(S\).

\begin{equation}
    h_{CLS}=H[0].
    \label{eq:cls-ff}
\end{equation}

Based on this feature aggregation, we get a global feature vector \(h_{CLS}\) containing multi-modal information. And then, the representation can be applied in classification layer. In our research we implement SoftMax function as output function.

\begin{equation}
    y=\text{SoftMax}\left(W\cdot h_{CLS}+b\right).
    \label{eq:softmax}
\end{equation}

\begin{figure}[H]
\centering
\includegraphics[width=\textwidth]{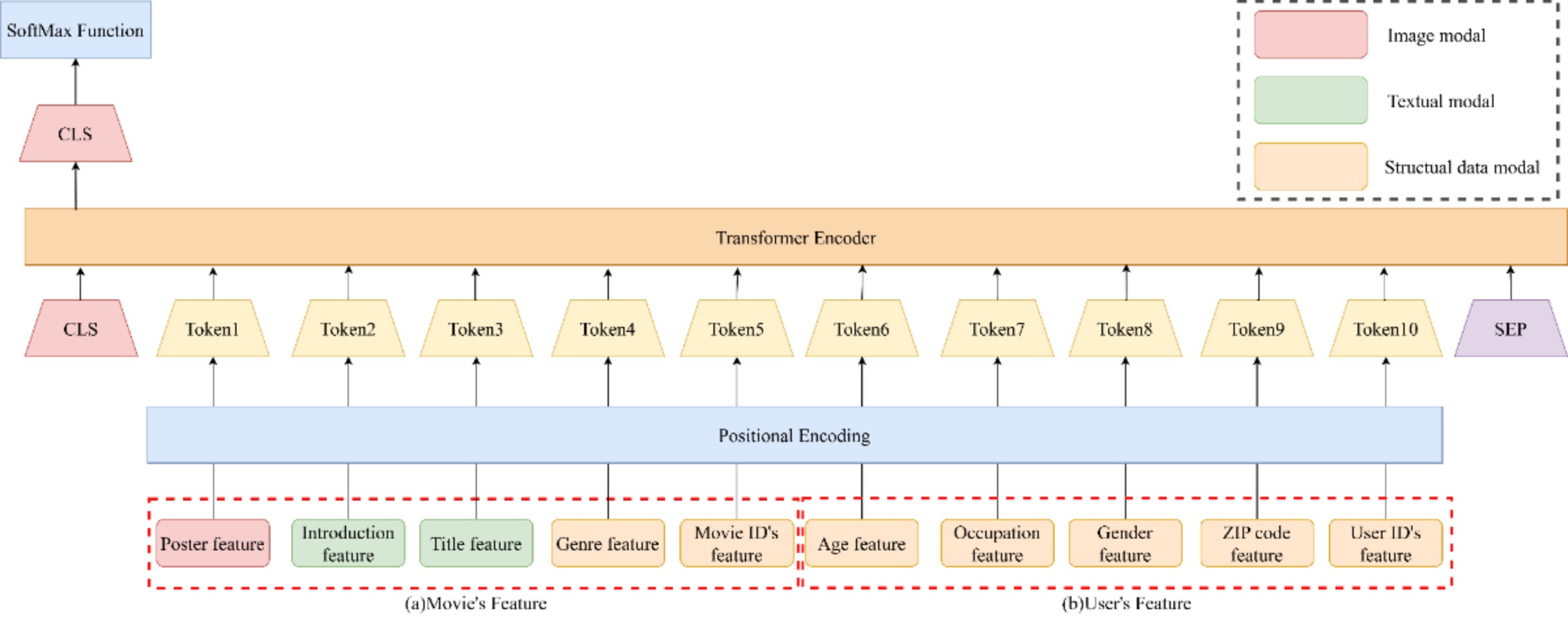}
\caption{Framework of proposed features fusion method, where \([CLS]\) and \([SEP]\) token is the beginning and the end of sequence. \label{fig:fusion}}
\end{figure}

The working process is shown in Figure \ref{fig:fusion}.

\section{Results}
In this section, we describe in detail the experiments we conducted to verify the validity of the model and test its performance. In this experiment, two datasets, MovieLens 1M and MovieLens 100k, were used as experimental data respectively to verify the validity of the model in different data. And we reproduce some classic algorithms such as item-CF, user-CF, and SVD. At the same time, we refer to the literature of Mu et al. \cite{b3} in multi-modal movie recommendation algorithm, and reproduce their work to compare with our proposed model.

\subsection{Hardware and Environment configure}

Our experiment used Google Colab as the computing platform, configured with an Nvidia A100 GPU and extended RAM. In the experimental environment configuration, the Python validation is 3.10.0, CUDA version is 12.1, Pytorch version is 2.3.0+cu121, torchvision version is 0.18.0+cu121, and transformers version is 4.30.0. In addition, the "Google-Bert/Bert-base-uncased" is used as the BERT model and the "google/ VIT-base-Patch16-224 "is used as the ViT model.

We split the MovieLens data set into a training set, a test set, and a verifier in a ratio of 9:0.5:0.5. We use the RMSE as an indicator to evaluate the predictive power of the model. The fomula of RMSE reads as:

\begin{equation}
\text{RMSE} = \sqrt{\frac{1}{N} \sum_{i=1}^{N} \left(y_i - \hat{y}_i\right)^2},
\label{eq:rmse}
\end{equation}where, \(N\) is number of samples and \(y_i\) is true label, \(\hat{y}_i\) is predicted label.

\subsection{Introduction of dataset}

The MovieLens dataset was collected and published by the GroupLens research group at the University of Minnesota to support research on movie recommender systems. This dataset is one of the widely used and studied data sets in the field of recommendation system, which contains users' ratings of movies and related movie information. The data sets used in this study are MovieLens 1M and MovieLens 100k. The MovieLens 100k dataset contains 100,000 movie ratings from 943 users and 1,682 movie ratings. Each user has reviewed at least five movies with ratings ranging from 1 to 5. The dataset also includes the year of release and genre of the film's title. Similar to MovieLens 100k The MovieLens 1M dataset contains the same content but the dataset is much larger, with 1 million ratings from more than 6,000 users and 3,900 movies. The MovieLens dataset covers all types of movies from all ages, and the diverse rating behaviors of users can be used to mine the preferences and rating patterns of different user groups.This research uses movie posters and movie introductions in movie data from IMDB datasets.

\begin{table}[]
    \centering
    \begin{tabular}{lcccc}
         \toprule
        \textbf{Dataset} & \textbf{Data} & \textbf{Value} & \textbf{Data size} & \textbf{Data size with poster} \\
        \midrule
        \multirow{4}{*}{MovieLens 100K} & Users & 943 &\multirow{4}{*}{5.3 MB} &\multirow{4}{*}{1.153 GB} \\
        & Movies & 1682 \\
        & Ratings & 100,000 \\
        & Sparsity & 93.695\% \\
        \midrule
        \multirow{4}{*}{MovieLens 1M} & Users & 6040  & \multirow{4}{*}{21.7 MB} & \multirow{4}{*}{11.530 GB} \\
        & Items & 3883 \\
        & Ratings & 1,000,209 \\
        & Sparsity & 95.359\% \\
        \bottomrule
    \end{tabular}
    \caption{Statistics of MovieLens datasets}
    \label{table:movielens}
\end{table}

In test A, we compare the proposed model with multiple baseline approaches includes three different categories: traditional method which used non-deep learning method and single modal information; deep learning method which used deep-learning technique but only single modal and multi-modal method which used deep-learning and multiple modals information.

For traditional method, we replicate User-CF, item-CF and SVD algorithm. For deep learning method we replicate Siet et al.'s \cite{Siet} work and Aljunid et al.'s \cite{Aljunid} work. 

For multi-modal method, we replicate the algorithm proposed by Mu et al. \cite{b3}, which use convolution neural network to embed the poster information and use dense layer to embed other information such as users' age, ZIP code movies' genre and so on. For poster data, we implement crawling technique to obtain movie poster and movie's introduction information from IMDB open source dataset in the data preparation phase. To evaluate the effect of multi-modal information, we divide the test into two different situations: single modal and cross modal, where single modal's input exclude poster and cross modal's input include poster. We evaluated the performance of different algorithms based on RMSE metrics. To verify the effectiveness of the cross-modal algorithms, we set up a control group without inputting movie poster information. The result is illustrated in Table \ref{tab:testA}.

\begin{table}[]
    \centering
    \begin{tabular}{cccc}
        \toprule
			\textbf{}	& \textbf{Approaches}	& \textbf{Train set}     & \textbf{Test set}\\
			\midrule
\multirow[m]{3}{*}{Traditional}	& Item-CF			& 2.534			& 2.812\\
			  	          & User-CF		  & 2.221		  & 2.310\\
                                & SVD			& 2.345			& 2.298\\
                                              \midrule
\multirow[m]{2}{*}{Deep learning}    & Siet's work			& 1.127			& 0.994\\
			  	                  & Aljunid's work			& 1.085			& 0.947\\
                   \midrule
\multirow[m]{4}{*}{Multi-modal}    & \(\text{Mu's work}_{sin}\)			& 1.137		& 1.004\\
			  	                 & \(\text{Mu's work}_{cro}\)		& 1.020		& 0.992\\
                                  & \(\text{Our work}_{sin}\)			& 1.065			& 0.987\\
			  	         & \(\text{Our work}_{cro}\)			& \textbf{0.965}			& \textbf{0.902}\\
			\bottomrule
    \end{tabular}
    \caption{Result of test A, where \(_{sin}\) stand for single modal and \(_{cro}\) stand for cross modal of multi-modal approaches. The test is based on MovieLens 100k dataset.}
    \label{tab:testA}
\end{table}

The following conclusions can be drawn from the results:

\begin{enumerate}
    \item Compared with the traditional models (item-CF, user-CF and SVD), the model proposed in this study has a large advantage in performance regardless of whether there is added image modality information or not.
    \item The advantages of the proposed model in feature extraction and feature fusion are verified compared to the method proposed by Mu et al.
    \item Inputting image information will improve the model performance to some extent compared to the model without image information.
\end{enumerate}

\subsection{Test B: explore the influence of size of dataset and learning rate}

In order to explore the performance in datasets of different sizes and the impact of different learning rates on model performance, we used the MovieLens 100k and MovieLens 1M datasets to train and evaluate the model proposed in this study under different learning rates. The learning rate is an important hyperparameter in deep learning, which can be pre-adjusted. A larger learning rate means the model will converge faster, but more unstable, and a smaller learning rate means slower convergence \cite{kane}. We set up the model's optimizer with a learning rate of 0.001, 0.0005 and 0.0001. The result is shown on Table \ref{tab:testB}.

\begin{table}[]
    \centering
    \begin{tabular}{ccccccc}
         \toprule
\textbf{} & \multicolumn{3}{c}{\textbf{MovieLens 100 K}} & \multicolumn{3}{c}{\textbf{MovieLens 1 M}} \\ 
\midrule
\textbf{Learning rate} & 0.001 &0.0005 &0.0001 & 0.001 &0.0005 &0.0001 \\ 
\textbf{RMSE of Train set} & 0.994 & \textbf{0.965} & 1.009 & 1.039 & 1.004 & 1.075 \\ 
\textbf{RMSE of Test set} & 0.991 & 0.902 & 1.058 & 0.996 & \textbf{0.893} & 0.994 \\ 
\bottomrule
    \end{tabular}
    \caption{RMSE scores of the multimodal movie recommendation algorithm based on deep learning.}
    \label{tab:testB}
\end{table}

At a learning rate of 0.0005, our proposed model performs 0.902 and 0.893 on the two datasets respectively, which are both slightly ahead of 0.0001 and 0.001. In addition, the model trained on a small dataset performs better on the training set, and the model trained on a large dataset performs better on the test set, which means that the generalization ability of the model will improve as the amount of data increases.

\section{Discussion and Conclusions}

In this research, we propose and evaluate a multi-modal information-based recommendation system. The proposed model extracts text modal information with BERT model and extracts image modal information with ViT model. It apply transformer architecture to fuse extracted feature and predicts users' rating. The main goal of this research is to verify the effectiveness of the recommendation system based on cross-modal information.

Experiments on the MovieLens 100k and 1M datasets show that our system outperforms traditional and deep learning baseline models. On MovieLens 100k, our system achieved an RMSE of 0.902, compared to 0.987 for a unimodal approach, indicating that multimodal data enhances prediction accuracy. Testing different learning rates (0.001, 0.0005, and 0.0001) revealed that 0.0005 yielded the best performance in both datasets. Additionally, as data volume increased, the system's generalization improved, achieving an RMSE of 0.893 on MovieLens 1M, demonstrating that more data enhances performance.

Despite the positive results of this study, there are still some limitations. Firstly, the large number of parameters in the pre-trained model and the high computational overhead in the tuning process may affect the economics of time-based recommendation in future applications, and secondly, the inclusion of multimodal data, especially image data, brings about noise enhancement. In future work, the focus should be on model simplification based on knowledge distillation and exploring techniques to deal with noise.


\section*{Declarations}

\begin{itemize}
\item Funding: This work was funded by the Natural Science Foundation of China (12271047); Guangdong Provincial Key Laboratory of Interdisciplinary Research and Application for Data Science, BNU-HKBU United International College (2022B1212010006); UIC research grant (R0400001-22; UICR0400008-21; UICR0400036-21CTL; UICR04202405-21); Guangdong College Enhancement and Innovation Program (2021ZDZX1046).
\item Conflict of interest: We declare that we have no financial and personal relationships with other
people or organizations that can inappropriately influence our work, and there is no professional or
other personal interests of any nature or kind in any product, service, and/or company that could be
construed as influencing the position presented in, or the review of, the manuscript entitled.
\item Data availability:  The experiment data is available in MovieLens official website (\url{https://movielens.org/}) and IMDB official website (\url{https://www.imdb.com/}), the data used in the experiment also can be downloaded through \url{https://github.com/Xia12121/MovieLens-dataset-with-poster}.
\item Author contribution: Conceptualization and methodology,L.X., software,L.X. and Y.Y., validation, Z.C. and Z.Y., writing--original draft, L.X., writing--review \& editing, S.Z., Z.C. and Z.Y., supervision, S.Z.
\end{itemize}


\bibliography{sn-bibliography}


\begin{thebibliography}{65}
\ifx \bisbn   \undefined \def \bisbn  #1{ISBN #1}\fi
\ifx \binits  \undefined \def \binits#1{#1}\fi
\ifx \bauthor  \undefined \def \bauthor#1{#1}\fi
\ifx \batitle  \undefined \def \batitle#1{#1}\fi
\ifx \bjtitle  \undefined \def \bjtitle#1{#1}\fi
\ifx \bvolume  \undefined \def \bvolume#1{\textbf{#1}}\fi
\ifx \byear  \undefined \def \byear#1{#1}\fi
\ifx \bissue  \undefined \def \bissue#1{#1}\fi
\ifx \bfpage  \undefined \def \bfpage#1{#1}\fi
\ifx \blpage  \undefined \def \blpage #1{#1}\fi
\ifx \burl  \undefined \def \burl#1{\textsf{#1}}\fi
\ifx \doiurl  \undefined \def \doiurl#1{\url{https://doi.org/#1}}\fi
\ifx \betal  \undefined \def \betal{\textit{et al.}}\fi
\ifx \binstitute  \undefined \def \binstitute#1{#1}\fi
\ifx \binstitutionaled  \undefined \def \binstitutionaled#1{#1}\fi
\ifx \bctitle  \undefined \def \bctitle#1{#1}\fi
\ifx \beditor  \undefined \def \beditor#1{#1}\fi
\ifx \bpublisher  \undefined \def \bpublisher#1{#1}\fi
\ifx \bbtitle  \undefined \def \bbtitle#1{#1}\fi
\ifx \bedition  \undefined \def \bedition#1{#1}\fi
\ifx \bseriesno  \undefined \def \bseriesno#1{#1}\fi
\ifx \blocation  \undefined \def \blocation#1{#1}\fi
\ifx \bsertitle  \undefined \def \bsertitle#1{#1}\fi
\ifx \bsnm \undefined \def \bsnm#1{#1}\fi
\ifx \bsuffix \undefined \def \bsuffix#1{#1}\fi
\ifx \bparticle \undefined \def \bparticle#1{#1}\fi
\ifx \barticle \undefined \def \barticle#1{#1}\fi
\bibcommenthead
\ifx \bconfdate \undefined \def \bconfdate #1{#1}\fi
\ifx \botherref \undefined \def \botherref #1{#1}\fi
\ifx \url \undefined \def \url#1{\textsf{#1}}\fi
\ifx \bchapter \undefined \def \bchapter#1{#1}\fi
\ifx \bbook \undefined \def \bbook#1{#1}\fi
\ifx \bcomment \undefined \def \bcomment#1{#1}\fi
\ifx \oauthor \undefined \def \oauthor#1{#1}\fi
\ifx \citeauthoryear \undefined \def \citeauthoryear#1{#1}\fi
\ifx \endbibitem  \undefined \def \endbibitem {}\fi
\ifx \bconflocation  \undefined \def \bconflocation#1{#1}\fi
\ifx \arxivurl  \undefined \def \arxivurl#1{\textsf{#1}}\fi
\csname PreBibitemsHook\endcsname

\bibitem[\protect\citeauthoryear{Eppler and Mengis}{2004}]{b1}
\begin{barticle}
\bauthor{\bsnm{Eppler}, \binits{M.J.}},
\bauthor{\bsnm{Mengis}, \binits{J.}}:
\batitle{The concept of information overload: A review of literature from organization science, accounting, marketing, mis, and related disciplines}.
\bjtitle{The Information Society}
\bvolume{20}(\bissue{5}),
\bfpage{325}--\blpage{344}
(\byear{2004})
\end{barticle}
\endbibitem

\bibitem[\protect\citeauthoryear{Roetzel}{2019}]{b2}
\begin{barticle}
\bauthor{\bsnm{Roetzel}, \binits{P.G.}}:
\batitle{Information overload in the information age: a review of the literature from business administration, business psychology, and related disciplines with a bibliometric approach and framework development}.
\bjtitle{Business Research}
\bvolume{12}(\bissue{2}),
\bfpage{479}--\blpage{522}
(\byear{2019})
\end{barticle}
\endbibitem

\bibitem[\protect\citeauthoryear{Mu and Wu}{2023}]{b3}
\begin{barticle}
\bauthor{\bsnm{Mu}, \binits{Y.}},
\bauthor{\bsnm{Wu}, \binits{Y.}}:
\batitle{Multi-modal movie recommendation system using deep learning}.
\bjtitle{Mathematics}
\bvolume{11}(\bissue{4}),
\bfpage{895}
(\byear{2023})
\end{barticle}
\endbibitem

\bibitem[\protect\citeauthoryear{Burke et~al.}{2011}]{b4}
\begin{barticle}
\bauthor{\bsnm{Burke}, \binits{R.}},
\bauthor{\bsnm{Felfernig}, \binits{A.}},
\bauthor{\bsnm{Göker}, \binits{M.H.}}:
\batitle{Recommender systems: An overview}.
\bjtitle{AI Magazine}
\bvolume{32}(\bissue{3}),
\bfpage{13}--\blpage{18}
(\byear{2011})
\end{barticle}
\endbibitem

\bibitem[\protect\citeauthoryear{Daneshvar and Ravanmehr}{2022}]{daneshvar}
\begin{barticle}
\bauthor{\bsnm{Daneshvar}, \binits{H.}},
\bauthor{\bsnm{Ravanmehr}, \binits{R.}}:
\batitle{A social hybrid recommendation system using lstm and cnn}.
\bjtitle{Concurrency and Computation: Practice and Experience}
\bvolume{34}(\bissue{18}),
\bfpage{7015}
(\byear{2022})
\end{barticle}
\endbibitem

\bibitem[\protect\citeauthoryear{Amatriain and Basilico}{2015}]{amatriain}
\begin{bchapter}
\bauthor{\bsnm{Amatriain}, \binits{X.}},
\bauthor{\bsnm{Basilico}, \binits{J.}}:
\bctitle{Recommender systems in industry: A netflix case study}.
In: \bbtitle{Recommender Systems Handbook},
pp. \bfpage{385}--\blpage{419}.
\bpublisher{Springer}, \blocation{???}
(\byear{2015})
\end{bchapter}
\endbibitem

\bibitem[\protect\citeauthoryear{Li et~al.}{2023}]{li}
\begin{barticle}
\bauthor{\bsnm{Li}, \binits{X.}},
\bauthor{\bsnm{Sun}, \binits{L.}},
\bauthor{\bsnm{Ling}, \binits{M.}},
\bauthor{\bsnm{Peng}, \binits{Y.}}:
\batitle{A survey of graph neural network based recommendation in social networks}.
\bjtitle{Neurocomputing}
\bvolume{549},
\bfpage{126441}
(\byear{2023})
\end{barticle}
\endbibitem

\bibitem[\protect\citeauthoryear{Zheng et~al.}{2024}]{zheng2024}
\begin{bchapter}
\bauthor{\bsnm{Zheng}, \binits{Y.}},
\bauthor{\bsnm{Lin}, \binits{X.}},
\bauthor{\bsnm{Chen}, \binits{K.}},
\bauthor{\bsnm{Zhu}, \binits{S.}}:
\bctitle{Cycletrans: a transformer-based clinical foundation model for safer prescription}.
In: \bbtitle{AAAI 2024 Spring Symposium on Clinical Foundation Models}
(\byear{2024})
\end{bchapter}
\endbibitem

\bibitem[\protect\citeauthoryear{Chen et~al.}{2020}]{chen2020}
\begin{barticle}
\bauthor{\bsnm{Chen}, \binits{Z.}},
\bauthor{\bsnm{Zhu}, \binits{S.}},
\bauthor{\bsnm{Niu}, \binits{Q.}},
\bauthor{\bsnm{Zuo}, \binits{T.}}:
\batitle{Knowledge discovery and recommendation with linear mixed model}.
\bjtitle{{IEEE Access}}
\bvolume{8},
\bfpage{38304}--\blpage{38317}
(\byear{2020})
\end{barticle}
\endbibitem

\bibitem[\protect\citeauthoryear{Lops et~al.}{2019}]{b5}
\begin{barticle}
\bauthor{\bsnm{Lops}, \binits{P.}}, \betal:
\batitle{Trends in content-based recommendation: Preface to the special issue on recommender systems based on rich item descriptions}.
\bjtitle{User Modeling and User-Adapted Interaction}
\bvolume{29},
\bfpage{239}--\blpage{249}
(\byear{2019})
\end{barticle}
\endbibitem

\bibitem[\protect\citeauthoryear{Chen et~al.}{2018}]{b6}
\begin{barticle}
\bauthor{\bsnm{Chen}, \binits{R.}}, \betal:
\batitle{A survey of collaborative filtering-based recommender systems: From traditional methods to hybrid methods based on social networks}.
\bjtitle{IEEE Access}
\bvolume{6},
\bfpage{64301}--\blpage{64320}
(\byear{2018})
\end{barticle}
\endbibitem

\bibitem[\protect\citeauthoryear{Çano and Morisio}{2017}]{b7}
\begin{barticle}
\bauthor{\bsnm{Çano}, \binits{E.}},
\bauthor{\bsnm{Morisio}, \binits{M.}}:
\batitle{Hybrid recommender systems: A systematic literature review}.
\bjtitle{Intelligent Data Analysis}
\bvolume{21}(\bissue{6}),
\bfpage{1487}--\blpage{1524}
(\byear{2017})
\end{barticle}
\endbibitem

\bibitem[\protect\citeauthoryear{Roy and Dutta}{2022}]{r1}
\begin{barticle}
\bauthor{\bsnm{Roy}, \binits{D.}},
\bauthor{\bsnm{Dutta}, \binits{M.}}:
\batitle{A systematic review and research perspective on recommender systems}.
\bjtitle{Journal of Big Data}
\bvolume{9}(\bissue{1}),
\bfpage{59}
(\byear{2022})
\end{barticle}
\endbibitem

\bibitem[\protect\citeauthoryear{Lu et~al.}{2019}]{lu2019}
\begin{bchapter}
\bauthor{\bsnm{Lu}, \binits{X.}},
\bauthor{\bsnm{Zhu}, \binits{S.}},
\bauthor{\bsnm{Niu}, \binits{Q.}},
\bauthor{\bsnm{Chen}, \binits{Z.}}:
\bctitle{Profile inference from heterogeneous data: Fundamentals and new trends}.
In: \bbtitle{International Conference on Business Information Systems},
pp. \bfpage{122}--\blpage{136}
(\byear{2019}).
\bcomment{Springer}
\end{bchapter}
\endbibitem

\bibitem[\protect\citeauthoryear{Dahdouh et~al.}{2019}]{dahdouh2019large}
\begin{barticle}
\bauthor{\bsnm{Dahdouh}, \binits{K.}},
\bauthor{\bsnm{Dakkak}, \binits{A.}},
\bauthor{\bsnm{Oughdir}, \binits{L.}},
\bauthor{\bsnm{Ibriz}, \binits{A.}}:
\batitle{Large-scale e-learning recommender system based on spark and hadoop}.
\bjtitle{Journal of Big Data}
\bvolume{6}(\bissue{1}),
\bfpage{1}--\blpage{23}
(\byear{2019})
\end{barticle}
\endbibitem

\bibitem[\protect\citeauthoryear{Achakulvisut et~al.}{2016}]{r3}
\begin{barticle}
\bauthor{\bsnm{Achakulvisut}, \binits{T.}}, \betal:
\batitle{Science concierge: A fast content-based recommendation system for scientific publications}.
\bjtitle{PloS One}
\bvolume{11}(\bissue{7}),
\bfpage{0158423}
(\byear{2016})
\end{barticle}
\endbibitem

\bibitem[\protect\citeauthoryear{Bergamaschi and Po}{2015}]{r4}
\begin{bchapter}
\bauthor{\bsnm{Bergamaschi}, \binits{S.}},
\bauthor{\bsnm{Po}, \binits{L.}}:
\bctitle{Comparing lda and lsa topic models for content-based movie recommendation systems}.
In: \bbtitle{Web Information Systems and Technologies: 10th International Conference, WEBIST 2014, Barcelona, Spain, April 3-5, 2014, Revised Selected Papers}
vol. \bseriesno{10},
(\byear{2015})
\end{bchapter}
\endbibitem

\bibitem[\protect\citeauthoryear{Bouihi and Bahaj}{2019}]{r5}
\begin{barticle}
\bauthor{\bsnm{Bouihi}, \binits{B.}},
\bauthor{\bsnm{Bahaj}, \binits{M.}}:
\batitle{Ontology and rule-based recommender system for e-learning applications}.
\bjtitle{International Journal of Emerging Technologies in Learning (Online)}
\bvolume{14}(\bissue{15}),
\bfpage{4}
(\byear{2019})
\end{barticle}
\endbibitem

\bibitem[\protect\citeauthoryear{Zhang and Iyengar}{2002}]{r6}
\begin{barticle}
\bauthor{\bsnm{Zhang}, \binits{T.}},
\bauthor{\bsnm{Iyengar}, \binits{V.S.}}:
\batitle{Recommender systems using linear classifiers}.
\bjtitle{The Journal of Machine Learning Research}
\bvolume{2},
\bfpage{313}--\blpage{334}
(\byear{2002})
\end{barticle}
\endbibitem

\bibitem[\protect\citeauthoryear{Gao et~al.}{2019}]{gao2019}
\begin{bchapter}
\bauthor{\bsnm{Gao}, \binits{B.}},
\bauthor{\bsnm{Zhan}, \binits{G.}},
\bauthor{\bsnm{Wang}, \binits{H.}},
\bauthor{\bsnm{Wang}, \binits{Y.}},
\bauthor{\bsnm{Zhu}, \binits{S.}}:
\bctitle{Learning with linear mixed model for group recommendation systems}.
In: \bbtitle{Proceedings of the 2019 11th International Conference on Machine Learning and Computing},
pp. \bfpage{81}--\blpage{85}
(\byear{2019})
\end{bchapter}
\endbibitem

\bibitem[\protect\citeauthoryear{Zuo et~al.}{2020}]{zuo2020}
\begin{bchapter}
\bauthor{\bsnm{Zuo}, \binits{T.}},
\bauthor{\bsnm{Zhu}, \binits{S.}},
\bauthor{\bsnm{Lu}, \binits{J.}}:
\bctitle{A hybrid recommender system combing singular value decomposition and linear mixed model}.
In: \bbtitle{Intelligent Computing. SAI 2020.}
vol. \bseriesno{1228},
pp. \bfpage{347}--\blpage{362}.
\bpublisher{Springer}, \blocation{???}
(\byear{2020})
\end{bchapter}
\endbibitem

\bibitem[\protect\citeauthoryear{Shafqat and Byun}{2020}]{r7}
\begin{barticle}
\bauthor{\bsnm{Shafqat}, \binits{W.}},
\bauthor{\bsnm{Byun}, \binits{Y.-C.}}:
\batitle{A context-aware location recommendation system for tourists using hierarchical lstm model}.
\bjtitle{Sustainability}
\bvolume{12}(\bissue{10}),
\bfpage{4107}
(\byear{2020})
\end{barticle}
\endbibitem

\bibitem[\protect\citeauthoryear{Hassan and Hamada}{2019}]{r9}
\begin{barticle}
\bauthor{\bsnm{Hassan}, \binits{M.}},
\bauthor{\bsnm{Hamada}, \binits{M.}}:
\batitle{Evaluating the performance of a neural network-based multi-criteria recommender system}.
\bjtitle{International Journal of Spatio-Temporal Data Science}
\bvolume{1}(\bissue{1}),
\bfpage{54}--\blpage{69}
(\byear{2019})
\end{barticle}
\endbibitem

\bibitem[\protect\citeauthoryear{Liu and Singh}{2016}]{r10}
\begin{bchapter}
\bauthor{\bsnm{Liu}, \binits{D.Z.}},
\bauthor{\bsnm{Singh}, \binits{G.}}:
\bctitle{A recurrent neural network based recommendation system}.
In: \bbtitle{International Conference on Recent Trends in Engineering, Science \& Technology}
(\byear{2016})
\end{bchapter}
\endbibitem

\bibitem[\protect\citeauthoryear{Huang et~al.}{2021}]{r11}
\begin{botherref}
\oauthor{\bsnm{Huang}, \binits{L.}}, et al.:
A deep reinforcement learning based long-term recommender system.
Knowledge-Based Systems
\textbf{213}
(2021)
\end{botherref}
\endbibitem

\bibitem[\protect\citeauthoryear{Lee et~al.}{2020}]{r12}
\begin{barticle}
\bauthor{\bsnm{Lee}, \binits{H.I.}}, \betal:
\batitle{A multi-period product recommender system in online food market based on recurrent neural networks}.
\bjtitle{Sustainability}
\bvolume{12}(\bissue{3}),
\bfpage{969}
(\byear{2020})
\end{barticle}
\endbibitem

\bibitem[\protect\citeauthoryear{Xu et~al.}{2021}]{xu2021}
\begin{bchapter}
\bauthor{\bsnm{Xu}, \binits{B.}},
\bauthor{\bsnm{Bu}, \binits{S.}},
\bauthor{\bsnm{Li}, \binits{X.}},
\bauthor{\bsnm{Lin}, \binits{Y.}},
\bauthor{\bsnm{Zhu}, \binits{S.}}:
\bctitle{xdeepfig: An extreme deep model with feature interactions and generation for ctr prediction}.
In: \bbtitle{Proceedings of the 2021 3rd International Conference on Big-data Service and Intelligent Computation},
pp. \bfpage{42}--\blpage{51}
(\byear{2021})
\end{bchapter}
\endbibitem

\bibitem[\protect\citeauthoryear{Feng et~al.}{2022}]{feng2022}
\begin{bchapter}
\bauthor{\bsnm{Feng}, \binits{Y.}},
\bauthor{\bsnm{Zhu}, \binits{S.}},
\bauthor{\bsnm{Ou}, \binits{Y.}}:
\bctitle{Accelerating din model for online ctr prediction with data compression}.
In: \bbtitle{2022 7th International Conference on Big Data Analytics (ICBDA)},
pp. \bfpage{84}--\blpage{89}
(\byear{2022}).
\bcomment{IEEE}
\end{bchapter}
\endbibitem

\bibitem[\protect\citeauthoryear{Luo et~al.}{2023}]{luo2023}
\begin{bchapter}
\bauthor{\bsnm{Luo}, \binits{Z.}},
\bauthor{\bsnm{Zhang}, \binits{Y.}},
\bauthor{\bsnm{Hu}, \binits{C.}},
\bauthor{\bsnm{Xia}, \binits{Y.}},
\bauthor{\bsnm{Zhu}, \binits{S.}}:
\bctitle{Click-through rate prediction models based on interest modeling}.
In: \bbtitle{Proceedings of the 2023 5th International Conference on Big Data Engineering},
pp. \bfpage{18}--\blpage{27}
(\byear{2023})
\end{bchapter}
\endbibitem

\bibitem[\protect\citeauthoryear{Siet et~al.}{2024}]{Siet}
\begin{botherref}
\oauthor{\bsnm{Siet}, \binits{S.}},
\oauthor{\bsnm{Peng}, \binits{S.}},
\oauthor{\bsnm{Ilkhomjon}, \binits{S.}},
\oauthor{\bsnm{Kang}, \binits{M.}},
\oauthor{\bsnm{Park}, \binits{D.-S.}}:
Enhancing sequence movie recommendation system using deep learning and kmeans.
Applied Sciences
\textbf{14}(6)
(2024)
\end{botherref}
\endbibitem

\bibitem[\protect\citeauthoryear{Aljunid and Dh}{2020}]{Aljunid}
\begin{barticle}
\bauthor{\bsnm{Aljunid}, \binits{M.F.}},
\bauthor{\bsnm{Dh}, \binits{M.}}:
\batitle{An efficient deep learning approach for collaborative filtering recommender system}.
\bjtitle{Procedia Computer Science}
\bvolume{171},
\bfpage{829}--\blpage{836}
(\byear{2020})
\end{barticle}
\endbibitem

\bibitem[\protect\citeauthoryear{Ma et~al.}{2015}]{r13}
\begin{barticle}
\bauthor{\bsnm{Ma}, \binits{T.}}, \betal:
\batitle{Social network and tag sources based augmenting collaborative recommender system}.
\bjtitle{IEICE transactions on Information and Systems}
\bvolume{98}(\bissue{4}),
\bfpage{902}--\blpage{910}
(\byear{2015})
\end{barticle}
\endbibitem

\bibitem[\protect\citeauthoryear{Millan et~al.}{2007}]{c1}
\begin{bchapter}
\bauthor{\bsnm{Millan}, \binits{M.}},
\bauthor{\bsnm{Trujillo}, \binits{M.}},
\bauthor{\bsnm{Ortiz}, \binits{E.}}:
\bctitle{A collaborative recommender system based on asymmetric user similarity}.
In: \bbtitle{Intelligent Data Engineering and Automated Learning-IDEAL 2007: 8th International Conference, Birmingham, UK, December 16-19, 2007. Proceedings}
vol. \bseriesno{8},
(\byear{2007})
\end{bchapter}
\endbibitem

\bibitem[\protect\citeauthoryear{O’connor et~al.}{2001}]{c2}
\begin{bchapter}
\bauthor{\bsnm{O’connor}, \binits{M.}},
\bauthor{\bsnm{Cosley}, \binits{D.}},
\bauthor{\bsnm{Konstan}, \binits{J.A.}},
\bauthor{\bsnm{Riedl}, \binits{J.}}:
\bctitle{Polylens: A recommender system for groups of users}.
In: \bbtitle{ECSCW 2001: Proceedings of the Seventh European Conference on Computer Supported Cooperative Work 16--20 September 2001, Bonn, Germany},
pp. \bfpage{199}--\blpage{218}
(\byear{2001}).
\bcomment{Springer}
\end{bchapter}
\endbibitem

\bibitem[\protect\citeauthoryear{Zhang et~al.}{}]{c3}
\begin{botherref}
\oauthor{\bsnm{Zhang}, \binits{C.}},
\oauthor{\bsnm{Yu}, \binits{L.}},
\oauthor{\bsnm{Wang}, \binits{Y.}},
\oauthor{\bsnm{Shah}, \binits{C.}},
\oauthor{\bsnm{Zhang}, \binits{X.}}:
Collaborative User Network Embedding for Social Recommender Systems,
pp. 381--389.
\doiurl{10.1137/1.9781611974973.43} .
\url{https://epubs.siam.org/doi/abs/10.1137/1.9781611974973.43}
\end{botherref}
\endbibitem

\bibitem[\protect\citeauthoryear{Fakhri et~al.}{2019}]{c4}
\begin{botherref}
\oauthor{\bsnm{Fakhri}, \binits{A.A.}},
\oauthor{\bsnm{Baizal}, \binits{Z.K.A.}},
\oauthor{\bsnm{Setiawan}, \binits{E.B.}}:
Restaurant recommender system using user-based collaborative filtering approach: a case study at bandung raya region.
Journal of Physics: Conference Series
\textbf{1192}(1)
(2019)
\end{botherref}
\endbibitem

\bibitem[\protect\citeauthoryear{Fu et~al.}{2018}]{c5}
\begin{barticle}
\bauthor{\bsnm{Fu}, \binits{M.}}, \betal:
\batitle{A novel deep learning-based collaborative filtering model for recommendation system}.
\bjtitle{IEEE Transactions on Cybernetics}
\bvolume{49}(\bissue{3}),
\bfpage{1084}--\blpage{1096}
(\byear{2018})
\end{barticle}
\endbibitem

\bibitem[\protect\citeauthoryear{Bobadilla et~al.}{2020}]{c6}
\begin{barticle}
\bauthor{\bsnm{Bobadilla}, \binits{J.}},
\bauthor{\bsnm{Alonso}, \binits{S.}},
\bauthor{\bsnm{Hernando}, \binits{A.}}:
\batitle{Deep learning architecture for collaborative filtering recommender systems}.
\bjtitle{Applied Sciences}
\bvolume{10}(\bissue{7}),
\bfpage{2441}
(\byear{2020})
\end{barticle}
\endbibitem

\bibitem[\protect\citeauthoryear{Wang et~al.}{2015}]{c7}
\begin{bchapter}
\bauthor{\bsnm{Wang}, \binits{H.}},
\bauthor{\bsnm{Wang}, \binits{N.}},
\bauthor{\bsnm{Yeung}, \binits{D.-Y.}}:
\bctitle{Collaborative deep learning for recommender systems}.
In: \bbtitle{Proceedings of the 21th ACM SIGKDD International Conference on Knowledge Discovery and Data Mining}
(\byear{2015})
\end{bchapter}
\endbibitem

\bibitem[\protect\citeauthoryear{Aljunid and Dh}{2020}]{c8}
\begin{barticle}
\bauthor{\bsnm{Aljunid}, \binits{M.F.}},
\bauthor{\bsnm{Dh}, \binits{M.}}:
\batitle{An efficient deep learning approach for collaborative filtering recommender system}.
\bjtitle{Procedia Computer Science}
\bvolume{171},
\bfpage{829}--\blpage{836}
(\byear{2020})
\end{barticle}
\endbibitem

\bibitem[\protect\citeauthoryear{Wu et~al.}{2022}]{c9}
\begin{barticle}
\bauthor{\bsnm{Wu}, \binits{S.}}, \betal:
\batitle{Graph neural networks in recommender systems: a survey}.
\bjtitle{ACM Computing Surveys}
\bvolume{55}(\bissue{5}),
\bfpage{1}--\blpage{37}
(\byear{2022})
\end{barticle}
\endbibitem

\bibitem[\protect\citeauthoryear{Wang et~al.}{2023}]{c11}
\begin{botherref}
\oauthor{\bsnm{Wang}, \binits{W.}}, et al.:
User-context collaboration and tensor factorization for gnn-based social recommendation.
IEEE Transactions on Network Science and Engineering
(2023)
\end{botherref}
\endbibitem

\bibitem[\protect\citeauthoryear{Kreutz and Schenkel}{2022}]{kreutz}
\begin{barticle}
\bauthor{\bsnm{Kreutz}, \binits{C.}},
\bauthor{\bsnm{Schenkel}, \binits{R.}}:
\batitle{Scientific paper recommendation systems: A literature review of recent publications}.
\bjtitle{International Journal of Digital Libraries}
\bvolume{23}(\bissue{4}),
\bfpage{335}--\blpage{369}
(\byear{2022})
\end{barticle}
\endbibitem

\bibitem[\protect\citeauthoryear{Pradhan and Pal}{2020}]{pradhan}
\begin{barticle}
\bauthor{\bsnm{Pradhan}, \binits{T.}},
\bauthor{\bsnm{Pal}, \binits{S.}}:
\batitle{A hybrid personalized scholarly venue recommender system integrating social network analysis and contextual similarity}.
\bjtitle{Future Generation Computer Systems}
\bvolume{110},
\bfpage{1139}--\blpage{1166}
(\byear{2020})
\end{barticle}
\endbibitem

\bibitem[\protect\citeauthoryear{Lorbeer et~al.}{2018}]{lorbeer}
\begin{barticle}
\bauthor{\bsnm{Lorbeer}, \binits{B.}}, \betal:
\batitle{A novel hybrid paper recommendation system using deep learning}.
\bjtitle{Big Data Research}
\bvolume{11},
\bfpage{44}--\blpage{53}
(\byear{2018})
\end{barticle}
\endbibitem

\bibitem[\protect\citeauthoryear{Qaiser and Ali}{2018}]{qaiser}
\begin{botherref}
\oauthor{\bsnm{Qaiser}, \binits{S.}},
\oauthor{\bsnm{Ali}, \binits{R.}}:
A scalable hybrid research paper recommender system for microsoft academic.
arXiv preprint arXiv:1905.08880
(2018)
\end{botherref}
\endbibitem

\bibitem[\protect\citeauthoryear{Biswas and Liu}{2021}]{biswas}
\begin{botherref}
\oauthor{\bsnm{Biswas}, \binits{P.K.}},
\oauthor{\bsnm{Liu}, \binits{S.}}:
A hybrid recommender system for recommending smartphones to prospective customers.
Papers With Code
(2021)
\end{botherref}
\endbibitem

\bibitem[\protect\citeauthoryear{Saga and Duan}{2018}]{c12}
\begin{bchapter}
\bauthor{\bsnm{Saga}, \binits{R.}},
\bauthor{\bsnm{Duan}, \binits{Y.}}:
\bctitle{Apparel goods recommender system based on image shape features extracted by a cnn}.
In: \bbtitle{2018 IEEE International Conference on Systems, Man, and Cybernetics (SMC)}
(\byear{2018})
\end{bchapter}
\endbibitem

\bibitem[\protect\citeauthoryear{Choudhury et~al.}{2021}]{m1}
\begin{barticle}
\bauthor{\bsnm{Choudhury}, \binits{S.S.}},
\bauthor{\bsnm{Mohanty}, \binits{S.N.}},
\bauthor{\bsnm{Jagadev}, \binits{A.K.}}:
\batitle{Multi-modal trust based recommender system with machine learning approaches for movie recommendation}.
\bjtitle{International Journal of Information Technology}
\bvolume{13},
\bfpage{475}--\blpage{482}
(\byear{2021})
\end{barticle}
\endbibitem

\bibitem[\protect\citeauthoryear{Nikzad-Khasmakhi et~al.}{2021}]{m2}
\begin{botherref}
\oauthor{\bsnm{Nikzad-Khasmakhi}, \binits{N.}}, et al.:
Berters: multi-modal representation learning for expert recommendation system with transformers and graph embeddings.
Chaos, Solitons \& Fractals
\textbf{151}
(2021)
\end{botherref}
\endbibitem

\bibitem[\protect\citeauthoryear{Chung and Chen}{2021}]{m3}
\begin{bchapter}
\bauthor{\bsnm{Chung}, \binits{Y.-H.}},
\bauthor{\bsnm{Chen}, \binits{Y.-L.}}:
\bctitle{Social recommendation system with multi-modal collaborative filtering}.
In: \bbtitle{2021 IEEE Global Communications Conference (GLOBECOM)}
(\byear{2021})
\end{bchapter}
\endbibitem

\bibitem[\protect\citeauthoryear{Sun et~al.}{2020}]{m4}
\begin{bchapter}
\bauthor{\bsnm{Sun}, \binits{R.}}, \betal:
\bctitle{Multi-modal knowledge graphs for recommender systems}.
In: \bbtitle{Proceedings of the 29th ACM International Conference on Information \& Knowledge Management}
(\byear{2020})
\end{bchapter}
\endbibitem

\bibitem[\protect\citeauthoryear{Atrey et~al.}{2010}]{m5}
\begin{barticle}
\bauthor{\bsnm{Atrey}, \binits{P.K.}}, \betal:
\batitle{Multi-modal fusion for multimedia analysis: a survey}.
\bjtitle{Multimedia systems}
\bvolume{16},
\bfpage{345}--\blpage{379}
(\byear{2010})
\end{barticle}
\endbibitem

\bibitem[\protect\citeauthoryear{Foresti and Snidaro}{2002}]{F1}
\begin{bchapter}
\bauthor{\bsnm{Foresti}, \binits{G.L.}},
\bauthor{\bsnm{Snidaro}, \binits{L.}}:
\bctitle{A distributed sensor network for video surveillance of outdoor environments}.
In: \bbtitle{Proceedings. International Conference on Image Processing},
vol. \bseriesno{1}
(\byear{2002})
\end{bchapter}
\endbibitem

\bibitem[\protect\citeauthoryear{Yang et~al.}{2005}]{F2}
\begin{barticle}
\bauthor{\bsnm{Yang}, \binits{M.-T.}},
\bauthor{\bsnm{Wang}, \binits{S.-C.}},
\bauthor{\bsnm{Lin}, \binits{Y.-Y.}}:
\batitle{A multi-modal fusion system for people detection and tracking}.
\bjtitle{International journal of imaging systems and technology}
\bvolume{15}(\bissue{2}),
\bfpage{131}--\blpage{142}
(\byear{2005})
\end{barticle}
\endbibitem

\bibitem[\protect\citeauthoryear{Zhu et~al.}{2020}]{F3}
\begin{barticle}
\bauthor{\bsnm{Zhu}, \binits{H.}}, \betal:
\batitle{Multi-modal fusion method based on self-attention mechanism}.
\bjtitle{Wireless Communications and Mobile Computing}
\bvolume{2020},
\bfpage{1}--\blpage{8}
(\byear{2020})
\end{barticle}
\endbibitem

\bibitem[\protect\citeauthoryear{Wei et~al.}{2020}]{F4}
\begin{bchapter}
\bauthor{\bsnm{Wei}, \binits{X.}}, \betal:
\bctitle{Multi-modality cross attention network for image and sentence matching}.
In: \bbtitle{Proceedings of the IEEE/CVF Conference on Computer Vision and Pattern Recognition}
(\byear{2020})
\end{bchapter}
\endbibitem

\bibitem[\protect\citeauthoryear{Wang et~al.}{2022}]{F5}
\begin{bchapter}
\bauthor{\bsnm{Wang}, \binits{Y.}}, \betal:
\bctitle{Multi-modal token fusion for vision transformers}.
In: \bbtitle{Proceedings of the IEEE/CVF Conference on Computer Vision and Pattern Recognition}
(\byear{2022})
\end{bchapter}
\endbibitem

\bibitem[\protect\citeauthoryear{Devlin et~al.}{2019}]{devlin}
\begin{bchapter}
\bauthor{\bsnm{Devlin}, \binits{J.}},
\bauthor{\bsnm{Chang}, \binits{M.-W.}},
\bauthor{\bsnm{Lee}, \binits{K.}},
\bauthor{\bsnm{Toutanova}, \binits{K.}}:
\bctitle{Bert: Pre-training of deep bidirectional transformers for language understanding}.
In: \bbtitle{Proceedings of the 2019 Conference of the North American Chapter of the Association for Computational Linguistics: Human Language Technologies, Volume 1 (Long and Short Papers)},
pp. \bfpage{4171}--\blpage{4186}
(\byear{2019})
\end{bchapter}
\endbibitem

\bibitem[\protect\citeauthoryear{Floridi and Chiriatti}{2020}]{P1}
\begin{barticle}
\bauthor{\bsnm{Floridi}, \binits{L.}},
\bauthor{\bsnm{Chiriatti}, \binits{M.}}:
\batitle{Gpt-3: Its nature, scope, limits, and consequences}.
\bjtitle{Minds and Machines}
\bvolume{30},
\bfpage{681}--\blpage{694}
(\byear{2020})
\end{barticle}
\endbibitem

\bibitem[\protect\citeauthoryear{Liu et~al.}{2021}]{liu2021robustly}
\begin{bchapter}
\bauthor{\bsnm{Liu}, \binits{Z.}},
\bauthor{\bsnm{Lin}, \binits{W.}},
\bauthor{\bsnm{Shi}, \binits{Y.}},
\bauthor{\bsnm{Zhao}, \binits{J.}}:
\bctitle{A robustly optimized bert pre-training approach with post-training}.
In: \bbtitle{China National Conference on Chinese Computational Linguistics},
pp. \bfpage{471}--\blpage{484}
(\byear{2021}).
\bcomment{Springer}
\end{bchapter}
\endbibitem

\bibitem[\protect\citeauthoryear{Simonyan and Zisserman}{2014}]{P3}
\begin{botherref}
\oauthor{\bsnm{Simonyan}, \binits{K.}},
\oauthor{\bsnm{Zisserman}, \binits{A.}}:
Very deep convolutional networks for large-scale image recognition.
arXiv preprint arXiv:1409.1556
(2014)
\end{botherref}
\endbibitem

\bibitem[\protect\citeauthoryear{He et~al.}{2016}]{P4}
\begin{bchapter}
\bauthor{\bsnm{He}, \binits{K.}}, \betal:
\bctitle{Deep residual learning for image recognition}.
In: \bbtitle{Proceedings of the IEEE Conference on Computer Vision and Pattern Recognition}
(\byear{2016})
\end{bchapter}
\endbibitem

\bibitem[\protect\citeauthoryear{Han et~al.}{2022}]{P5}
\begin{barticle}
\bauthor{\bsnm{Han}, \binits{K.}}, \betal:
\batitle{A survey on vision transformer}.
\bjtitle{IEEE transactions on pattern analysis and machine intelligence}
\bvolume{45}(\bissue{1}),
\bfpage{87}--\blpage{110}
(\byear{2022})
\end{barticle}
\endbibitem

\bibitem[\protect\citeauthoryear{Kane}{2018}]{kane}
\begin{bbook}
\bauthor{\bsnm{Kane}, \binits{F.}}:
\bbtitle{Building Recommender Systems with Machine Learning and AI: Help People Discover New Products and Content with Deep Learning, Neural Networks, and Machine Learning Recommendations.}
\bpublisher{Independently published}, \blocation{???}
(\byear{2018})
\end{bbook}
\endbibitem

\end{thebibliography}

\end{document}